\documentclass{osa-article}

\journal{osac}
\usepackage{subfigure}  
\usepackage{color}

\articletype{Research Article}

\begin{document}

\title{Coupling between  subwavelength nano-slits lattice modes  and  metal-insulator-graphene cavity modes: A semi-analytical model}

\author{Kofi Edee\authormark{1}, Maha Benrhouma\authormark{1}, Mauro Antezza\authormark{2,3}, Jonathan Albert  Fan\authormark{4}, Brahim Guizal\authormark{2}}

\address{\authormark{1,*} Universit\'{e} Clermont Auvergne,
Institut Pascal, BP 10448, F-63000 Clermont-Ferrand, France, CNRS, UMR 6602, Institut Pascal, F-63177 Aubi\`{e}re, France\\
\authormark{2} Laboratoire Charles Coulomb (L2C), UMR 5221 CNRS-Universit\'{e} de Montpellier, F-34095 Montpellier, France\\
\authormark{3}Institut Universitaire de France, 1 rue Descartes, F-75231 Paris Cedex 05, France\\
\authormark{4}Department of Electrical Engineering, Stanford University, Stanford, California 94305, United States}
\email{\authormark{*}kofi.edee@uca.fr} 

\ociscodes{(050.0050) Diffraction and gratings; (050.6624) Subwavelength structures.}

\begin{abstract}
We present a semi-analytical model of the resonance phenomena occurring in a hybrid system made of a 1D array of periodic subwavelength  slits deposited on an insulator/graphene layer. We show  that the spectral response of this hybrid system can be fully explained by a simple semi-analytical model based on a weak and strong couplings between two elementary sub-systems. The first elementary sub-system consists of a 1D array of periodic subwavelength  slits viewed as a homogeneous medium. In this medium lives a metal-insulator-metal lattice mode interacting with surface and cavity plasmon modes. A weak coupling with surface plasmon modes on both faces of the perforated metal film leads to a broadband spectrum while a strong coupling between this first sub-system and a second one made of a graphene-insulator-metal gap leads to a narrow band spectrum. We provide a semi-analytical model based on these two interactions allowing to efficiently access the full spectrum of the hybrid system.
\end{abstract}

\section{introduction}
Extraordinary optical transmission (EOT) \cite{Ebbesen1} through an opaque metallic film perforated with subwavelength slits has received great interest over the past decade because of its numerous applications in optoelectronics such as mid-infrared spatial light modulators, linear signal processing or biosensing.
Many theoretical and experimental works  were carried out in order  to understand and predict EOT and, especially, to highlight the role of surface waves \cite{Moreno,Aigouy,Haito1,Nikittin}. 
More recently we provided, in \cite{Kofi6},  a simple and versatile model, for this phenomenon, involving a specific mode living in an equivalent homogeneous medium and a phase correction to account for surface waves. The  proposed semi-analytical model is valid  from the visible to the infrared frequencies ranges. 
On the other hand, significant efforts have been made to create active or tunable plasmonic devices operating from THz to mid-infrared frequencies. Thanks to its extraordinary electronic and optical  properties, graphene, a single layer of arranged carbon atoms has attracted much attention in the last years. This material can support both TE an TM surface plasmons and can exhibit some remarkable properties such as flexible wide band tunability  that can be exploited to build new plasmonic devices. The main challenge when designing a graphene-plasmon-based device is how to efficiently excite graphene surface plasmons with a free space electromagnetic wave since there is a huge momentum mismatch  between the two electromagnetic modes. Generally two strategies are used. The first one consists in patterning the graphene sheet into nano-resonators \cite{Nikittin2, Thongrattanasiri,Yan, Rodrigo, Brar, Strait, Yi, Koppens, Yan2, Fang, Fallahi, Zhao,Amin}. In this case a surface plasmon of the obtained structure which is very similar to the graphene surface plasmon is excited and an absorption rate close to $100\%$ can be reached. In particular in \cite{Amin}, the authors presented an electrically tunable hybrid graphene-gold Fano resonator which  consists of a square graphene patch and a square gold frame. They showed that the  destructive interference between the narrow- and broadband dipolar surface plasmons, which are induced respectively on the surfaces of the graphene patch and the gold frame, leads to the plasmonic equivalent of electromagnetically induced transparency (EIT). However  patterning a graphene sheet requires sophisticated processing techniques and deteriorates its extraordinary mobility.
The second strategy consists in using a continuous graphene sheet instead of undesirable patterned graphene structure \cite{Tang, Zizhuo, XZhao, Xia, Gao,Zhang}. In this approach, the graphene sheet is coupled with nano-scatterers  such as nano-particles, or nano-gratings. Gao {\it et al.} proposed \cite{Gao} to use diffractive gratings to create a guided-wave resonance in the graphene film that can be directly observed from the normal incidence transmission spectra. In \cite{XZhao} Zhao {\it et al.} studied a tunable plasmon-induced-transparency effect in a grating-coupled double-layer graphene hybrid system at far-infrared frequencies. They used a diffractive grating to couple a normal incident wave and plasmonic modes living in a system of two graphene-films separated by a spacer. Zhang {\it et al.} \cite{Zhang} investigated optical field enhancements, in a wide mid-infrared band, originating from the excitation of graphene plasmons, by introducing a dielectric grating underneath a graphene monolayer. 
Usually,  the optical response of all the grating-graphene based structures listed above is performed thanks to the finite difference time domain method (FDTD) or to the finite element method (FEM). However the features of these hybrid graphene-resonators devices is often linked to a plasmon resonance phenomenon. Therefore a modal method allowing for a full modal analysis of the couplings occurring in these plasmonic systems seems more suitable.\\
In this paper, we investigate an optical tunable plasmonic system involving two fundamental phenomena: an EOT phenomenon and a metal-insulator-graphene cavity plasmon mode excitation. We propose a semi-analytical model allowing to fully describe the spectrum behaviour of an hybrid plasmonic structure, made of a 1D periodic subwavelength slits array deposited on an insulator/graphene layers. The spectrum of the proposed hybrid system exhibits Lorentz and Fano-like resonances and also  other  broadband and narrow band resonances that are efficiently captured by  our simplified model. In order to explain the origin of this particular behaviour, we first split the hybrid system into a couple of sub-systems. Second, thanks to a modal analysis through the polynomial modal method (PMM: one of the most efficient methods for modeling the electromagnetic properties of periodic structures) \cite{Kofi1,Kofi2,Kofi3,Kofi4}, we demonstrate that the scattering parameters of each sub-system can be computed through a concept of weak and strong couplings. Finally we provide analytical expressions of the reflection and transmission coefficients of the structure and describe the mechanisms leading to Lorentz and Fano resonances occurring in it.    
\section{Physical system}
\begin{figure}
\centering 
{\includegraphics[width=0.6\textwidth]{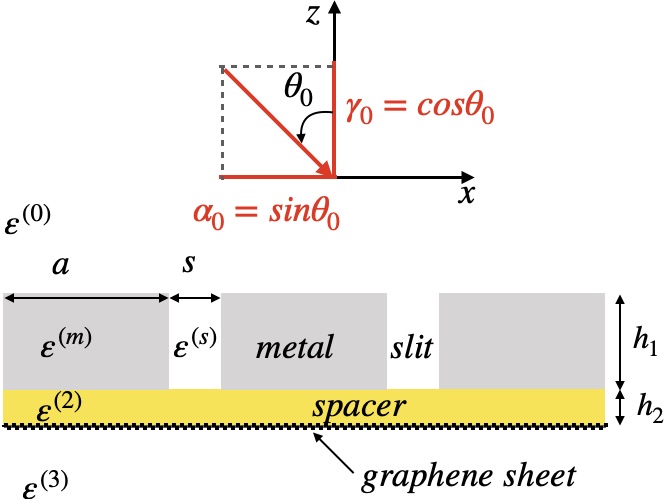}}
\caption{\label{geometry} Sketch of hybrid structure made of  a dispersive metal film perforated with a subwavelength periodic  array of 1D nano-slits deposited on a  dielectric spacer ended by a continuous graphene sheet.}
\end{figure}

\begin{figure}
\centering  
\subfigure [\label{spectrum_muc_1000_sp_10nm} $\mu_c=1eV$]
{\includegraphics[width=0.49\textwidth]{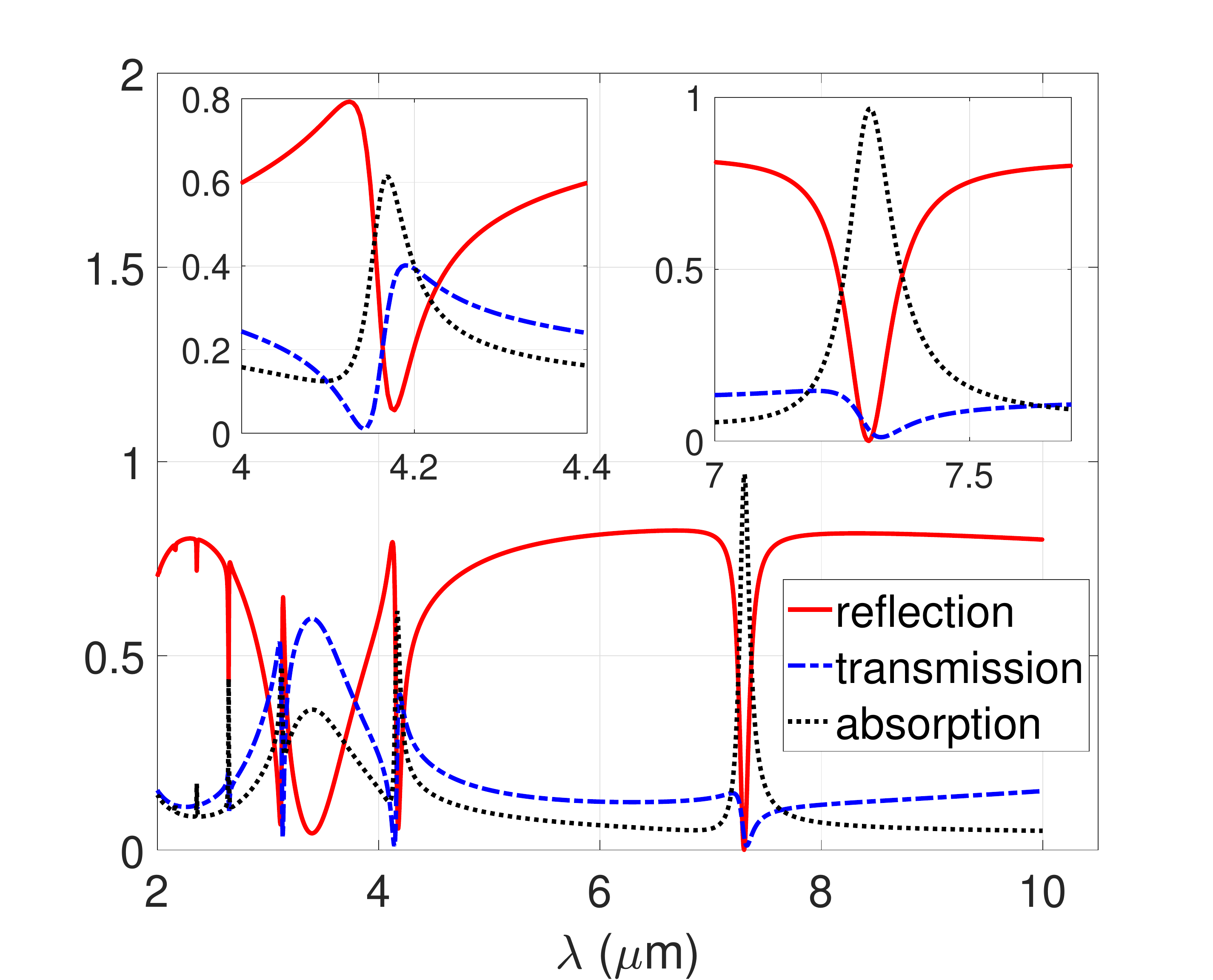}}
\centering  
\subfigure [\label{spectrum_muc_1500_sp_10nm}$\mu_c=1.5eV$]
{\includegraphics[width=0.49\textwidth]{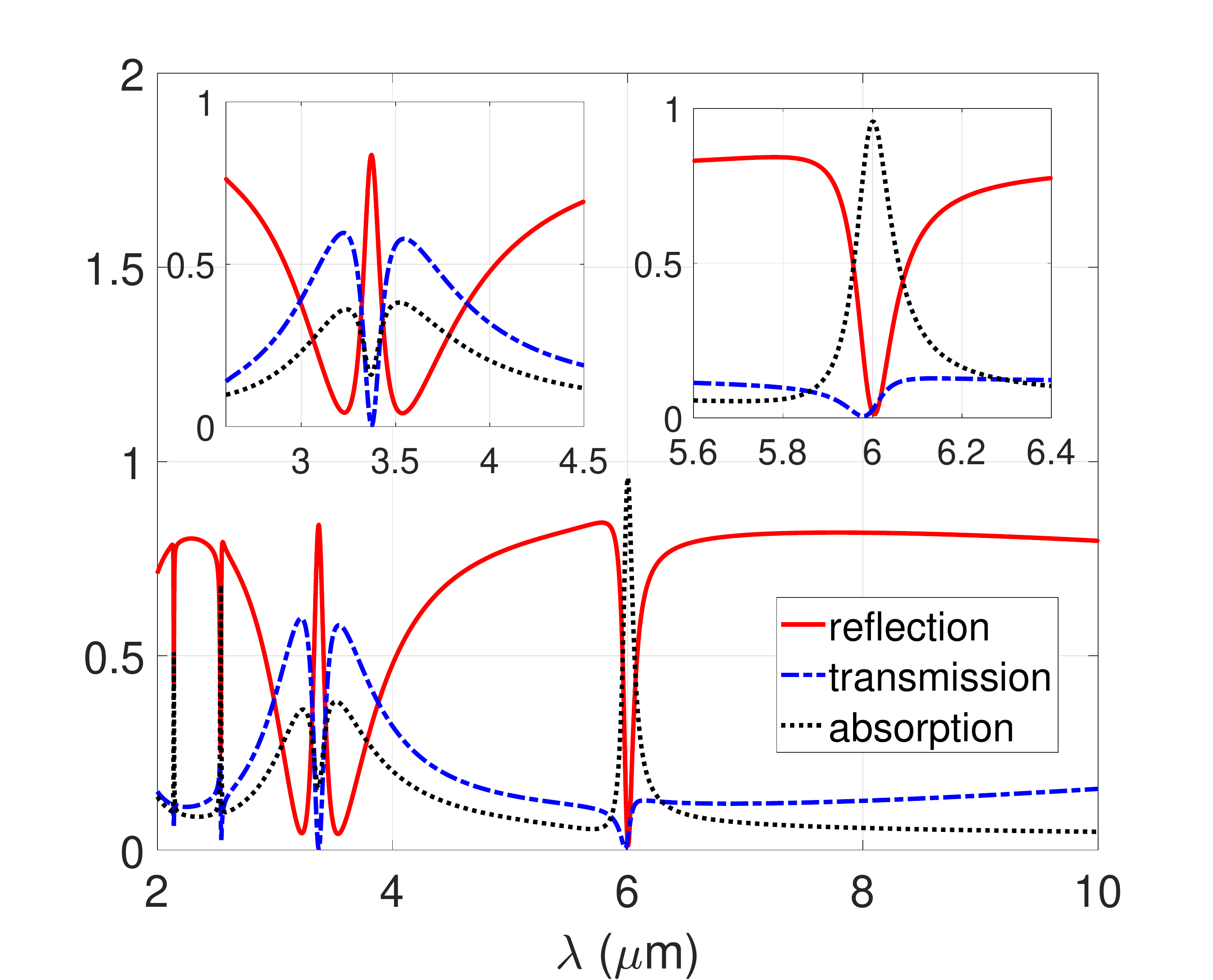}}
\caption{Reflection, transmission and absorption spectra of the hybrid system for $\mu_c=1eV$ (Fig. (\ref{spectrum_muc_1000_sp_10nm})) and $\mu_c=1.5eV$ (Fig. (\ref{spectrum_muc_1500_sp_10nm})). The hybrid structure exhibits both broadband and tunable narrow band resonances with respect to the chemical potential. Parameters: $\varepsilon^{1}=\varepsilon^{3}=\varepsilon^{slit}=1$, incidence angle= $0^{o}$, $h=800nm$, $d=165nm$, $a=15nm$.}
\end{figure}

\begin{figure}
\centering  
\subfigure [\label{real_Hy_muc_1000_lambda_417_10nm} $\lambda=4.17 \mu m$]
{\includegraphics[width=0.49\textwidth]{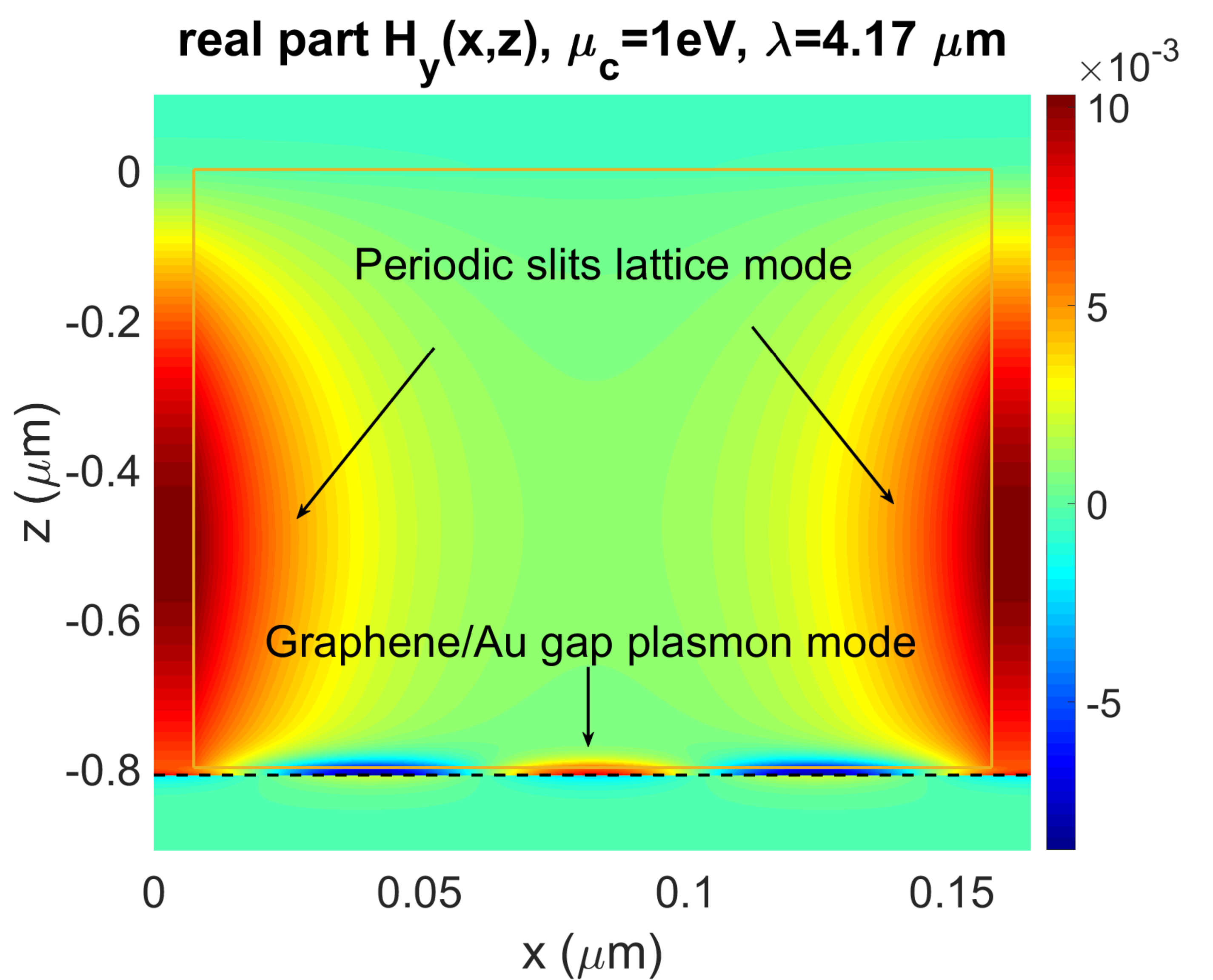}}
\centering  
\subfigure [\label{real_Hy_muc_1000_lambda_730_10nm} $\lambda=7.30 \mu m$]
{\includegraphics[width=0.49\textwidth]{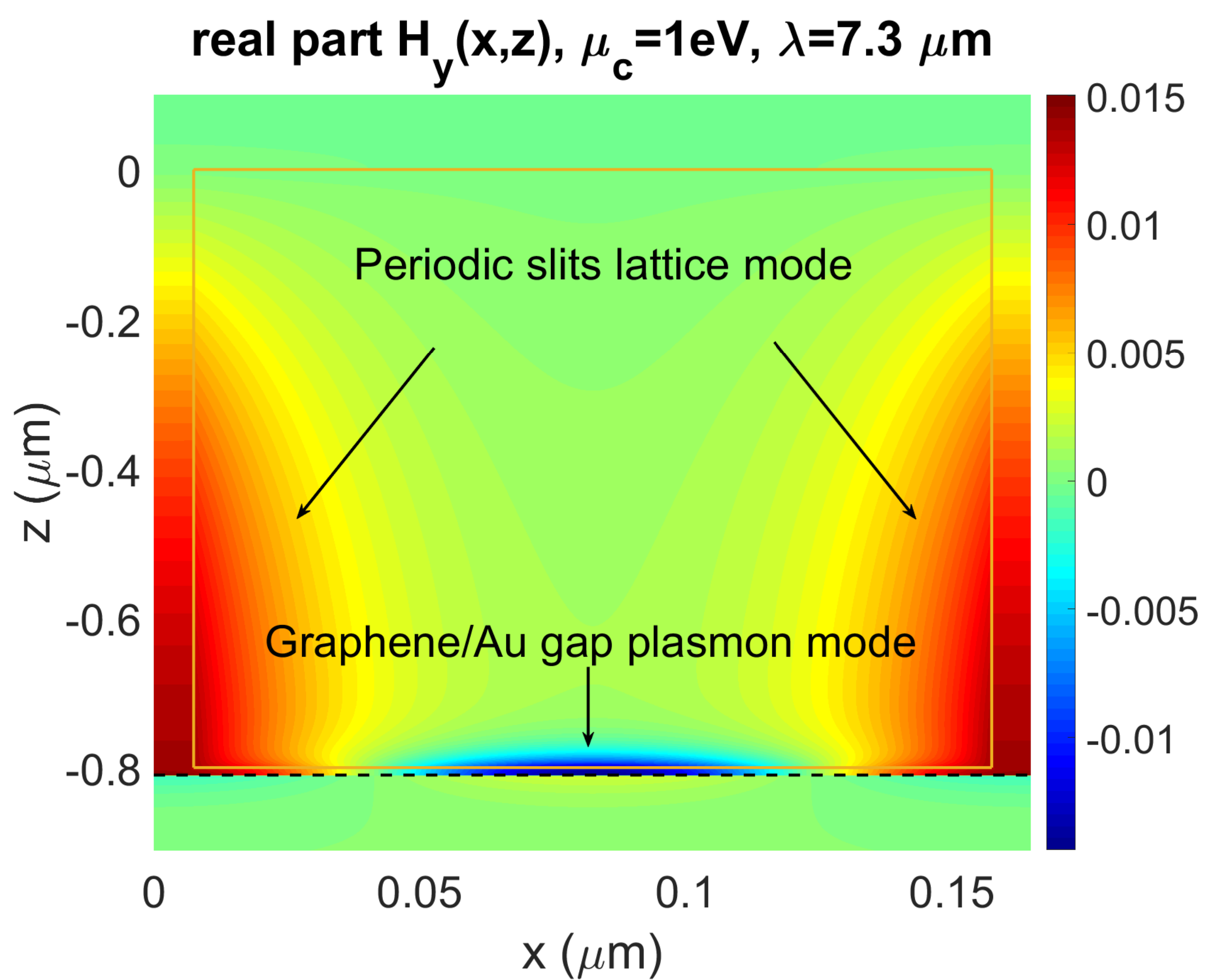}}
\caption{Real part  of the magnetic field $H_x(x,z)$ at $\lambda=4.17 \mu m$ (Fig. (\ref{real_Hy_muc_1000_lambda_417_10nm})) and at $\lambda=7.30 \mu m$ (Fig. (\ref{real_Hy_muc_1000_lambda_730_10nm})). Parameters: $\varepsilon^{1}=\varepsilon^{3}=\varepsilon^{slit}=1$, incidence angle= $0^{o}$, $h=800nm$, $d=165nm$, $a=15nm$.}
\end{figure}

The hybrid structure under study is presented in Fig. (\ref{geometry}). It consists of two sub-systems. The first sub-system earlier studied in \cite{Kofi6} is a sub-wavelength periodic array of nano-slits with  height $h_1=800nm$, period $d=165 nm << \lambda$ and slits-width $s=15nm$.  The relative permittivity of the material filling the slits is denoted by $\varepsilon^{(s)}$ while the dispersive relative permittivity of the metal (gold) is denoted by $\varepsilon^{(m)}$ and described by the Drude-Lorentz model \cite{BB, Rakic}. See reference \cite{Kofi6} for the numerical parameters used for $\varepsilon^{(m)}$ description. This first sub-structure is deposited on a dielectric spacer (with relative permittivity $\varepsilon^{(2)}=1.54^2$ and hight $h_2=10nm$) itself deposited on a continuous graphene sheet. 
The monolayer graphene optical properties are modeled with an equivalent layer with thickness $\Delta$ and permittivity $\varepsilon(\omega)$ \cite{Vakil} : 
\begin{equation}
\varepsilon(\omega)=1+i\dfrac{\sigma(\omega)}{\varepsilon_0 \omega \Delta}
\end{equation}
where the optical conductivity of graphene 
\begin{equation}\label{cond graphene}
\sigma(\omega)=\sigma_{inter}+\sigma_{intra}
\end{equation}
includes both the interband and intraband contributions.
The first term of Eq. (\ref{cond graphene}) \textsl{i.e.} the interband contribution $\sigma_{inter}$ has the form $\sigma_{inter}=\sigma_{inter}^{\prime}+i\sigma_{inter}^{\prime\prime}$, where
\begin{equation}
\left\{\begin{array}{l}
\sigma_{inter}^{\prime}=\sigma_0 \left[1+\dfrac{1}{\pi}atan\left(\dfrac{\hbar \omega -2\mu_c}{\hbar\Gamma}\right)-\dfrac{1}{\pi} atan\left(\dfrac{\hbar\omega +2\mu_c}{\hbar\Gamma}\right)\right]\\
\sigma_{inter}^{\prime\prime}=\dfrac{\sigma_0}{2\pi}ln\left[\dfrac{2\mu_c-\hbar\omega}{2\mu_c+\hbar \omega}\right].
\end{array}\right.
\end{equation}
$\sigma_0=\pi e^2/2h$ is the universal conductivity of the graphene, $1/\Gamma$ is the  relaxation time {{(throughout this work, we will take $\Gamma=2.10^{12}$s$^{-1}$)}} and $\mu_c$ is the Fermi level. 
The second  term $\sigma_{intra}$ of Eq. (\ref{cond graphene}) describes a Drude model response for intraband processes: 
\begin{equation}
\sigma_{intra}=\sigma_0\dfrac{4\mu_c}{\pi}\dfrac{1}{\hbar\Gamma-i\hbar\omega}.    
\end{equation}
This hybrid structure is excited, from the upper medium (having relative permittivity $\varepsilon^{(0)}$) by a TM polarized plane wave (the magnetic field is parallel to the $y$ axis). The  wave vector of the incident wave is denoted by  $\mathbf{K_0}=k_0\left(\alpha_0\mathbf{e_x}+\beta_0\mathbf{e_y}+\gamma_0\mathbf{e_z}\right)$, where $k_0=2\pi/\lambda=\omega/c$ denotes the wavenumber,  $\lambda$ being the wavelength and $c$ the light velocity in vacuum. The relative permittivity of the lower region is denoted by $\varepsilon^{(3)}$. 
We report in Figs. (\ref{spectrum_muc_1000_sp_10nm}) and (\ref{spectrum_muc_1500_sp_10nm}), the spectra of the hybrid structure for two values of the chemical potential: $\mu_c=1eV$ Fig. (\ref{spectrum_muc_1000_sp_10nm}) and $\mu_c=1.5eV$ Fig. (\ref{spectrum_muc_1500_sp_10nm}). These curves  display both broadband and narrow bands resonance phenomena. It has been shown in \cite{Kofi6} that a Lorentz-like resonance  corresponding to an EOT phenomenon  can occur in the first sub-system \textsl{i.e.} the dispersive metal film perforated with a subwavelength periodic  array of 1D nano-slits excited by a plane.  In the current case, this EOT occurs around $\lambda=3.37 \mu m$ and as pointed out in \cite{Kofi6} it is related to the excitation of a particular eigenmode of the slit grating structure : the so-called lattice mode. One can easily conceive that the broadband resonance is related to the EOT phenomenon outlined later, while the narrow band resonance phenomena are due to  Fabry-Perrot-like resonances of a cavity mode living in the metal/spacer/graphene gap. For example, for $\mu_c=1eV$, a first two narrow resonances are observed around $\lambda=4.17\mu m$ and $\lambda=7.3\mu m$.
The real parts of the magnetic field plotted  in Fig. (\ref{real_Hy_muc_1000_lambda_417_10nm}) at $\lambda=4.17\mu m$ and in Fig. (\ref{real_Hy_muc_1000_lambda_730_10nm}) at $\lambda=7.30 \mu m$, for $\mu_c=1eV$, support the fact that the narrow band resonances  are linked to the resonance of a cavity mode of the horizontal metal/insulator/graphene sub-system.
As the effective index of this mode strongly depends on the chemical potential $\mu_c$, the resonance frequencies of this hybrid cavity mode shift with increasing $\mu_c$. 
Comparing  the  reflection spectrum of the first sub-system with that of the hybrid structure, we can interpret the latter spectral response as a weak or strong coupling between the lattice mode of the former sub-system with the cavity mode of the metal/insulator/graphene gap. We propose in the following, a simple single mode model allowing to  efficiently describe, and understand the mechanism of this vertical-to-horizontal cavity modes coupling.  
\section{Modal analysis of the system}
\begin{figure}[!ht]
\centering 
{\includegraphics[width=0.7\textwidth]{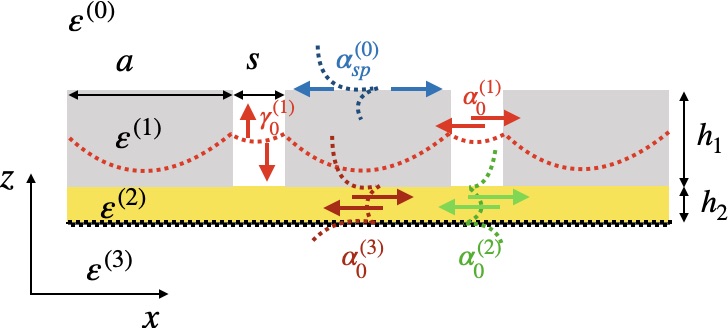}}
\caption{\label{structure2} Sketch of the mechanism of the coupling between cavity lattice modes of the periodic array of nano-slits and the metal/insulator/graphene gap plasmon modes. Strong and weak couplings between three modes are responsible of the resonance phenomena of the hybrid structure.}
\end{figure}
\begin{figure}[!ht]
\centering 
{\includegraphics[width=0.8\textwidth]{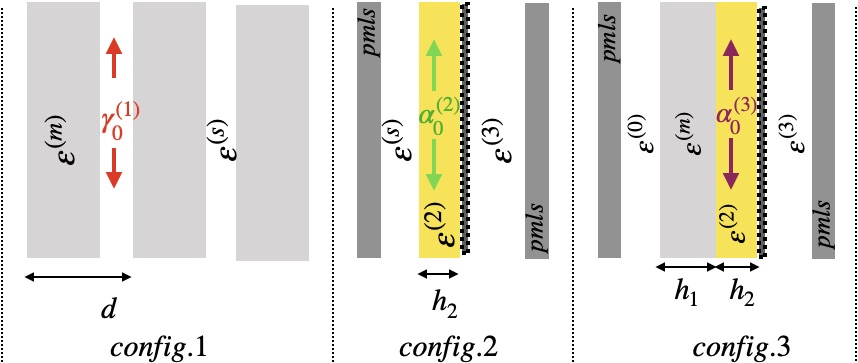}}
\caption{\label{structure7} Configurations used for the computation of the required effective indices (eigenvalues of equations \ref{op_TM}). $config.1$ is used for  the computation of the modes of periodic arrays of nano-slits in general and in particular for the computation of the cavity lattice mode effective index $\gamma_0^{(1)}$. $config.2$ is used for the computation of the effective index $\alpha_0^{(2)}$ of the  plasmon mode. The gap plasmon mode effective index $\alpha_0^{(3)}$ is computed thanks to $config.3$.}
\end{figure}
\begin{figure}[!ht]
\centering 
{\includegraphics[width=0.6\textwidth]{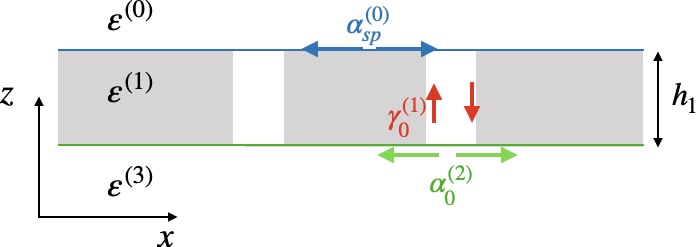}}
\caption{\label{sub_system_1a} Sketch of the weak coupling sub-system consisting of a periodic array of nano-slits encapsulated between $\varepsilon^{(0)}$ and $\varepsilon^{(3)}$ media. The lattice mode $\gamma_0^{(1)}$ is assumed to live in an $\sqrt{\varepsilon^{(1)}}$ effective homogeneous medium. Two plasmon modes $\alpha_{sp}^{(0)}$ and $\alpha_0^{(2)}$ ensure the phase matching with the plane waves in media $\varepsilon^{(0)}$ and $\varepsilon^{(3)}$.}
\end{figure}
\begin{figure}[!ht]
\centering 
{\includegraphics[width=0.6\textwidth]{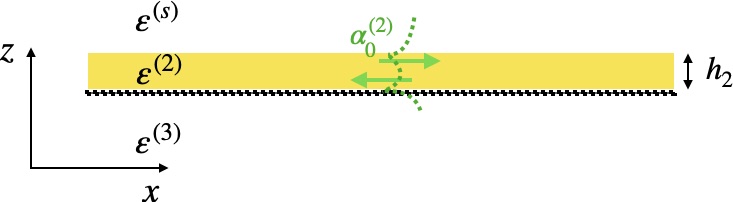}}
\caption{\label{sub_system_1b} The sketch of $\alpha_0^{(2)}$ plasmon mode computation.}
\end{figure}

The sketch of vertical-to-horizontal cavity modes coupling outlined in the previous section is presented in Fig. (\ref{structure2}), where $\gamma_0^{(1)}$  denotes effective index of the periodic slits array lattice mode in the $z$-direction while $\alpha_0^{(3)}$ denotes that of the metal/insulator/graphene cavity mode in the $x$-direction. The effective indices $\alpha_0^{(1)}$ and $\alpha_0^{(2)}$ will be introduced later. Modal methods are very suitable to deal with the current problem since it is related to mode resonances. Thus, all required effective indices are computed as eigenvalues of the generic operator $\mathcal{L}^{(k)}$:
\begin{equation} \label{op_TM}
\mathcal{L}^{(k)}(\omega)|H^{(k)}_{q}(\omega)\rangle =(\gamma_q^{(k)}(\omega))^2 |H^{(k)}_{q}(\omega) \rangle\\ 
\end{equation} 
with
\begin{equation*} \label{vap_vep}
\mathcal{L}^{(k)}(x,\omega)=\left(\dfrac{c}{\omega}\right)^2 \varepsilon^{(k)}(x,\omega) \partial_x \frac{1}{\varepsilon^{(k)}(x,\omega)}\partial_x+ \varepsilon^{(k)}(x,\omega).
\end{equation*} 
Figure \ref{structure7} illustrates the different configurations used for the computation of the required effective indices. Recall that these effective indices are computed as eigenvalues of equation (\ref{op_TM}). The first configuration ($config.1$) is used for the computation of the modes of periodic arrays of nano-slits in general and particularly for the computation of the cavity lattice mode effective index $\gamma_0^{(1)}$. The second configuration ($config.2$) is used for the computation of the plasmon mode effective index $\alpha_0^{(2)}$ while the cavity plasmon mode effective index $\alpha_0^{(3)}$ is computed thanks to the third configuration ($config.3$). Practically, the PMM is used to solve numerically the eigenvalue equation Eq. (\ref{op_TM}).
For that purpose, the structure is divided  into sub-intervals $I^{(k)}_x$, in the $x$-direction: $k\in\{1,2\}$ for $config.1$ while $k\in\{1,6\}$ for $config.2$  and $config.3$.
At this stage, we split the hybrid system into two coupled sub-systems:
\begin{itemize}
\item a weakly coupled sub-system sketched in figures \ref{sub_system_1a} and \ref{sub_system_1b} which leads to the broadband resonances. 
\item a strongly coupled sub-system sketched in figures \ref{sub_system_2a} and \ref{sub_system_2b} leading to a narrow bands dispersion curves.
\end{itemize}
Let us now analyse each coupled sub-system and provide semi-analytical models allowing to describe them. 

\subsection{Weakly coupled sub-system}
\begin{figure}
\centering  
\subfigure [\label{compare_reflec_12_muc_1000_sp_10nm} weakly coupled sub-system response]
{\includegraphics[width=0.49\textwidth]{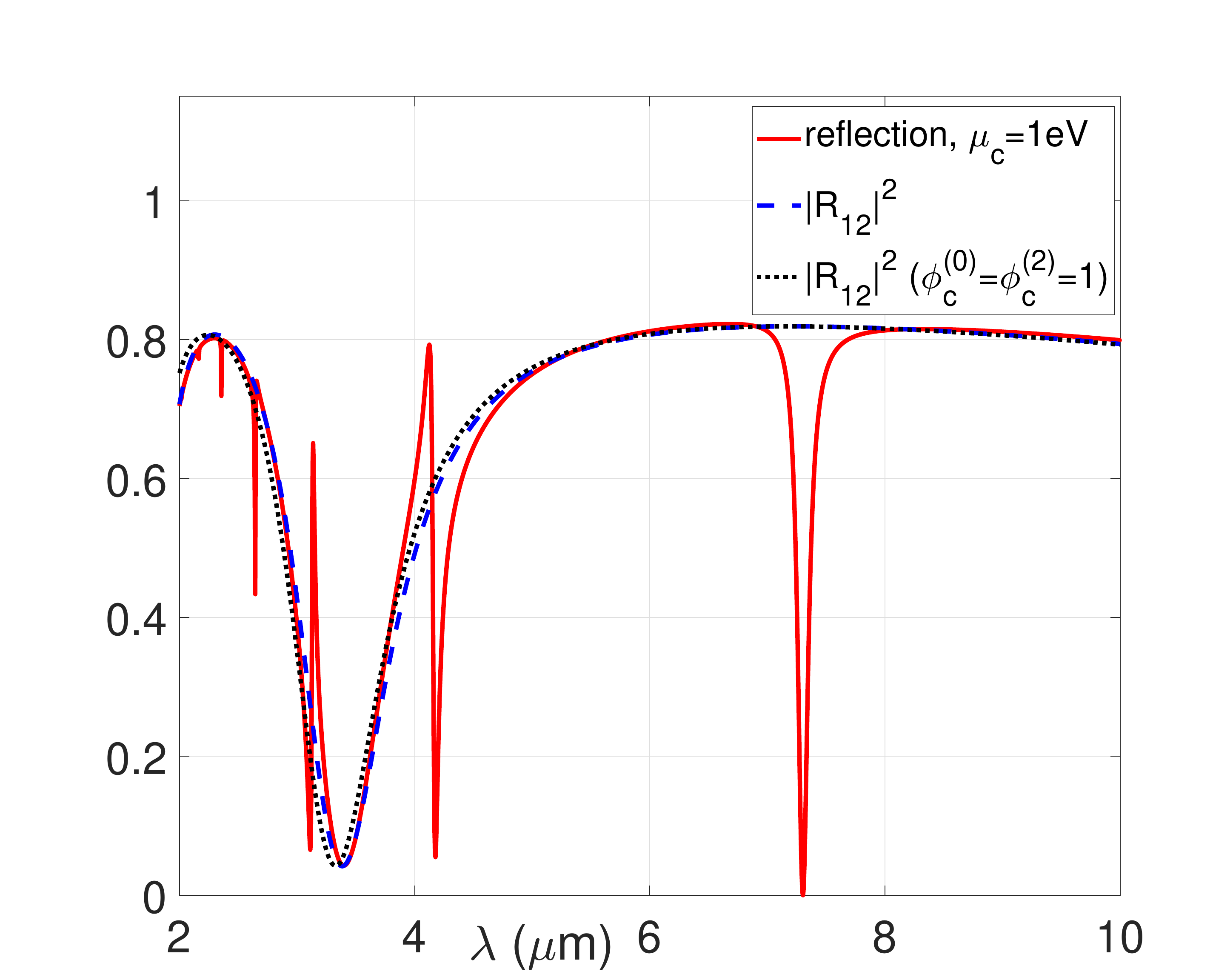}}
\centering  
\subfigure [\label{compare_reflec_13_muc_1000_sp_10nm} strongly coupled sub-system response]
{\includegraphics[width=0.49\textwidth]{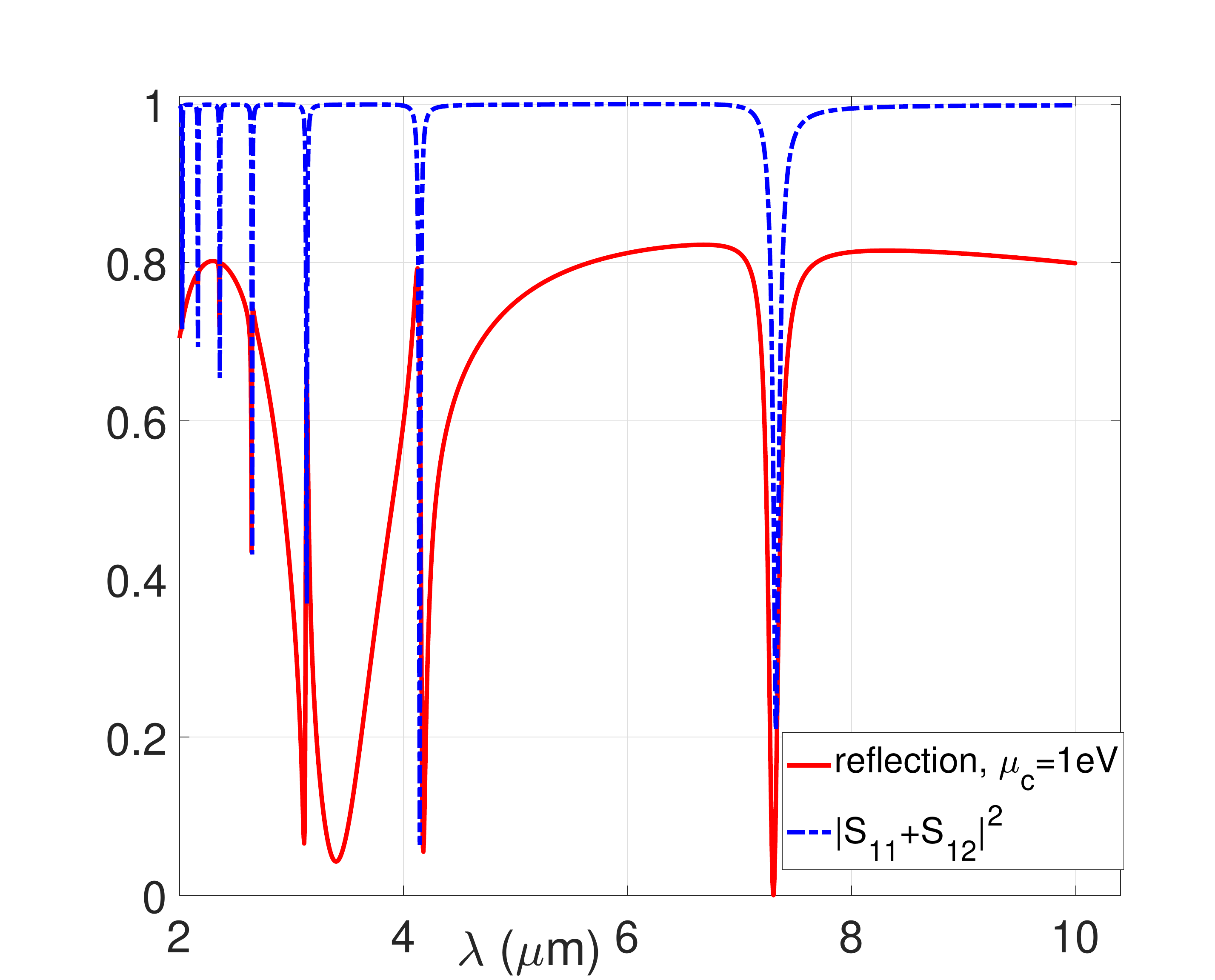}}
\caption{Comparison between the reflection spectrum of the hybrid structure and the responses  of the weakly coupled sub-system (a) and the strongly coupled sub-system (b). As expected, the weakly coupled sub-system reflection spectrum $|R_{12}(\lambda)|^2$ perfectly matches the broadband resonances of the hybrid structure. On the other hand, the strongly coupled sub-system spectrum characteristic function $|S_{11}(\lambda)+S_{12}(\lambda)|^2$ perfectly matches the narrow band resonances  of the hybrid structure. Parameters:  $\lambda \in [2,10]\mu m$, $\varepsilon^{(1)}=\varepsilon^{(3)}=\varepsilon^{(s)}=1$, $\varepsilon^{(2)}=1.54^2$, incidence angle= $0^{o}$, $\mu_c=1eV$.}
\end{figure}

A semi-analytical model for the weakly coupled system has been already described in \cite{Kofi6}. This system consists of a periodic array of subwavelength nano-slits encapsulated between media with relative permittivities $\varepsilon^{(0)}$ and $\varepsilon^{(3)}$. As pointed out in \cite{Kofi6}, the electromagnetic response of the system to an incident plane wave excitation, in the static limit ($d<<\lambda$), is  equivalent to that of a slab with equivalent permittivity $\varepsilon^{(1)}=\langle 1/\varepsilon^{(m,s)}(x) \rangle ^{-1}$ and height $h_1$.   
Its reflection and transmission  coefficients $R_{12}$ and $T_{12}$ are then given by :
\begin{equation}\label{reflection}
R_{12}=\dfrac{r_1+\phi_1 r_2 \phi_2}{1+r_1\phi_1 r_2 \phi_2}
\end{equation}
and
\begin{equation}\label{transmission}
T_{12}=\dfrac{t_1t_2 \phi_2}{1+r_1\phi_1 r_2 \phi_2}
\end{equation}
where  $r_{1}$, $t_{1}$ and $r_{2}$, $t_{2}$  are the Fresnel coefficients at the interfaces $\varepsilon^{(0)}/\varepsilon^{(1)}$  and $\varepsilon^{(1)}/\varepsilon^{(3)}$ under TM polarization:
\begin{eqnarray}
r_1=\dfrac{1-n_{01(\omega)}}{1+n_{01(\omega)}},\quad r_2=\dfrac{1-n_{13(\omega)}}{1+n_{13(\omega)}}, \\
t_1=\dfrac{2}{1+n_{01(\omega)}},\quad t_2=\dfrac{2}{1+n_{13(\omega)}}.
\end{eqnarray}
where
\begin{equation}
n_{01}(\omega)=\dfrac{\gamma_0^{(1)}(\omega)/\varepsilon^{(1)}(\omega)}{\gamma_0^{(0)}(\omega)/\varepsilon^{(0)}(\omega)}, \quad \text{ and } \quad n_{13}(\omega)=\dfrac{\gamma_0^{(3)}(\omega)/\varepsilon^{(3)}(\omega)}{\gamma_0^{(1)}(\omega)/\varepsilon^{(1)}(\omega)},
\end{equation}
and
\begin{eqnarray}\label{phase }
\phi_1=e^{-i k_0\gamma^{(1)}_0 h_1}\phi_c^{(0)}, \quad
\phi_2=e^{-i k_0\gamma^{(1)}_0 h_1}\phi_c^{(2)}
\end{eqnarray}
with
\begin{eqnarray}\label{phase correction}
\phi_c^{(0)}=e^{-i k_0\alpha_{sp}^{(0)} a^{(0)}}, \quad
\phi_c^{(2)}=e^{-i k_0\alpha_{0}^{(2)} a^{(2)}}.
\end{eqnarray}
Phase correction terms are introduced in order to take into account the phase matching between the lattice mode with effective index $\gamma_0^{(1)}$ and the incident plane wave (see \cite{Kofi6}). In equation Eq. (\ref{phase correction}), $\alpha_{sp}^{(0)}=\sqrt{\dfrac{{\varepsilon^{(0)}\varepsilon^{(m)}}}{\varepsilon^{(0)}+\varepsilon^{(m)}}}$ is the effective index of the surface plasmon propagating along the  upper interface, $a^{(1)}=\dfrac{a}{4}\sqrt{\dfrac{\varepsilon^{(0)}}{\varepsilon^{(s)}}}$ and $a^{(2)}=\dfrac{a}{4}\sqrt{\dfrac{\varepsilon^{(3)}}{\varepsilon^{(s)}}}$. We compare  in Fig. (\ref{compare_reflec_12_muc_1000_sp_10nm}) the spectrum of the reflectivity $|R_{12}|^2$, with the reflectivity of the hybrid system. As expected, the $|R_{12}|^2$ curve perfectly matches the broadband resonance of the hybrid structure.
The impact of the phase correction terms  $\phi_c^{(0),(2)}$ on the results is not significant since the omission of these terms only induces a little shift of the  $|R_{12}|^2$ curve. This is why we consider  the coupling between the  $\gamma_0^{(1)}$-effective index-slit-mode and the $\alpha_0^{(2)}$-effective index-plasmon-mode as a weak coupling. 
\subsection{Strongly coupled system}
\begin{figure}[!ht]
\centering 
{\includegraphics[width=0.6\textwidth]{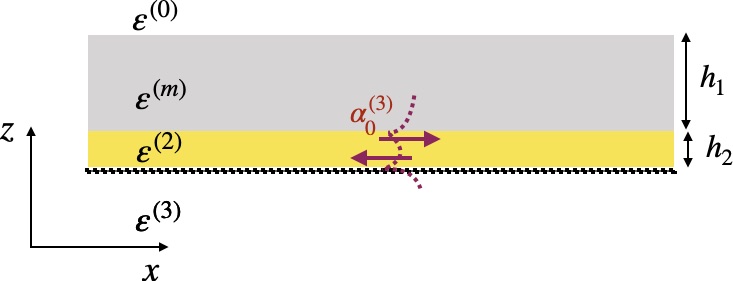}}
\caption{\label{sub_system_2b} The sketch of $\alpha_0^{(3)}$ plasmon mode computation.}
\end{figure}

\begin{figure}[ht]
\centering  
\subfigure [\label{compare_reflec_R_muc_1000_sp_10nm} Reflection for $\mu_c=1eV$]
{\includegraphics[width=0.49\textwidth]{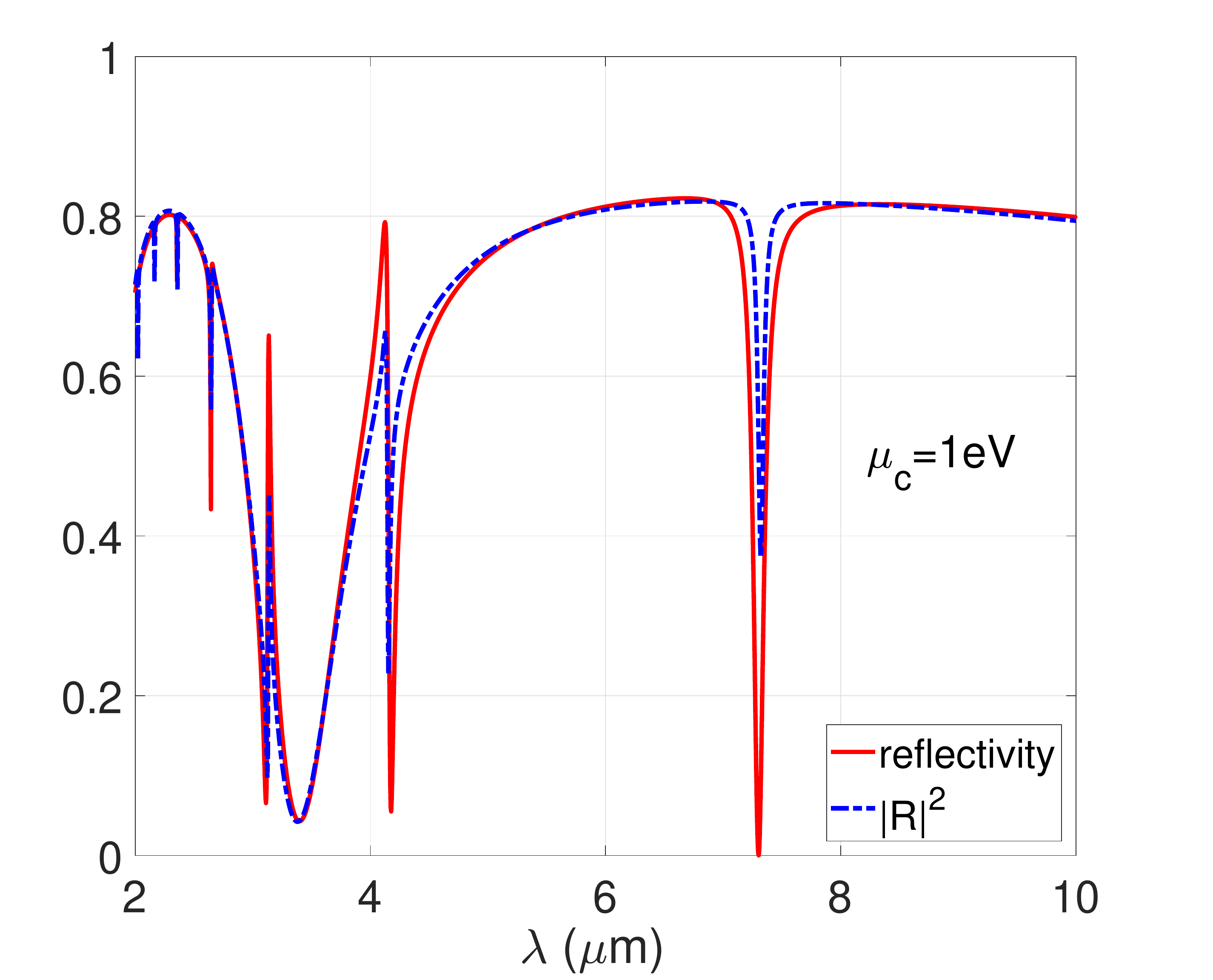}}
\centering  
\subfigure [\label{compare_trans_T_muc_1000_sp_10nm} transmission for $\mu_c=1eV$]
{\includegraphics[width=0.49\textwidth]{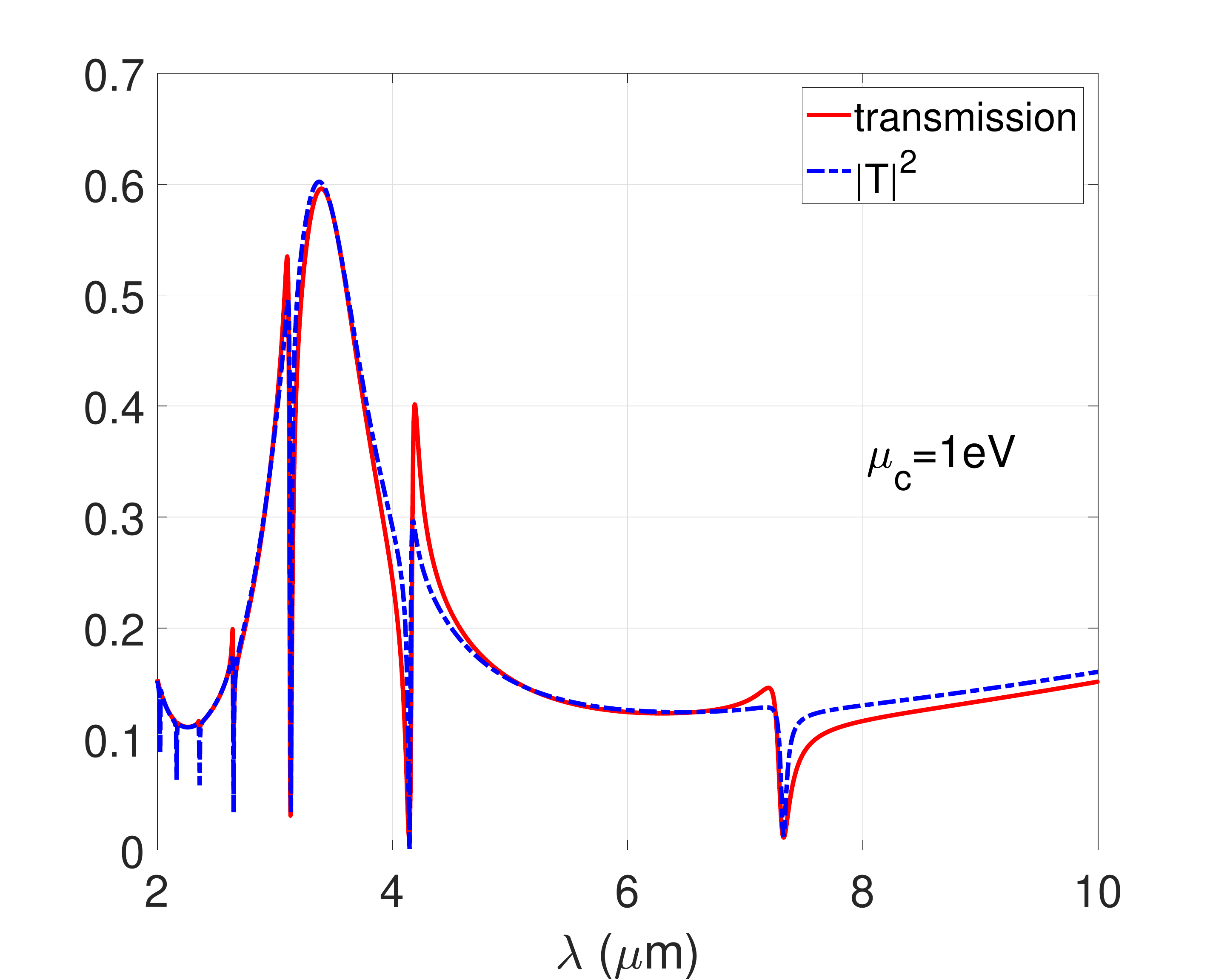}}
\subfigure [\label{compare_reflec_R_muc_1500_sp_10nm} Reflection for $\mu_c=1.5eV$]
{\includegraphics[width=0.49\textwidth]{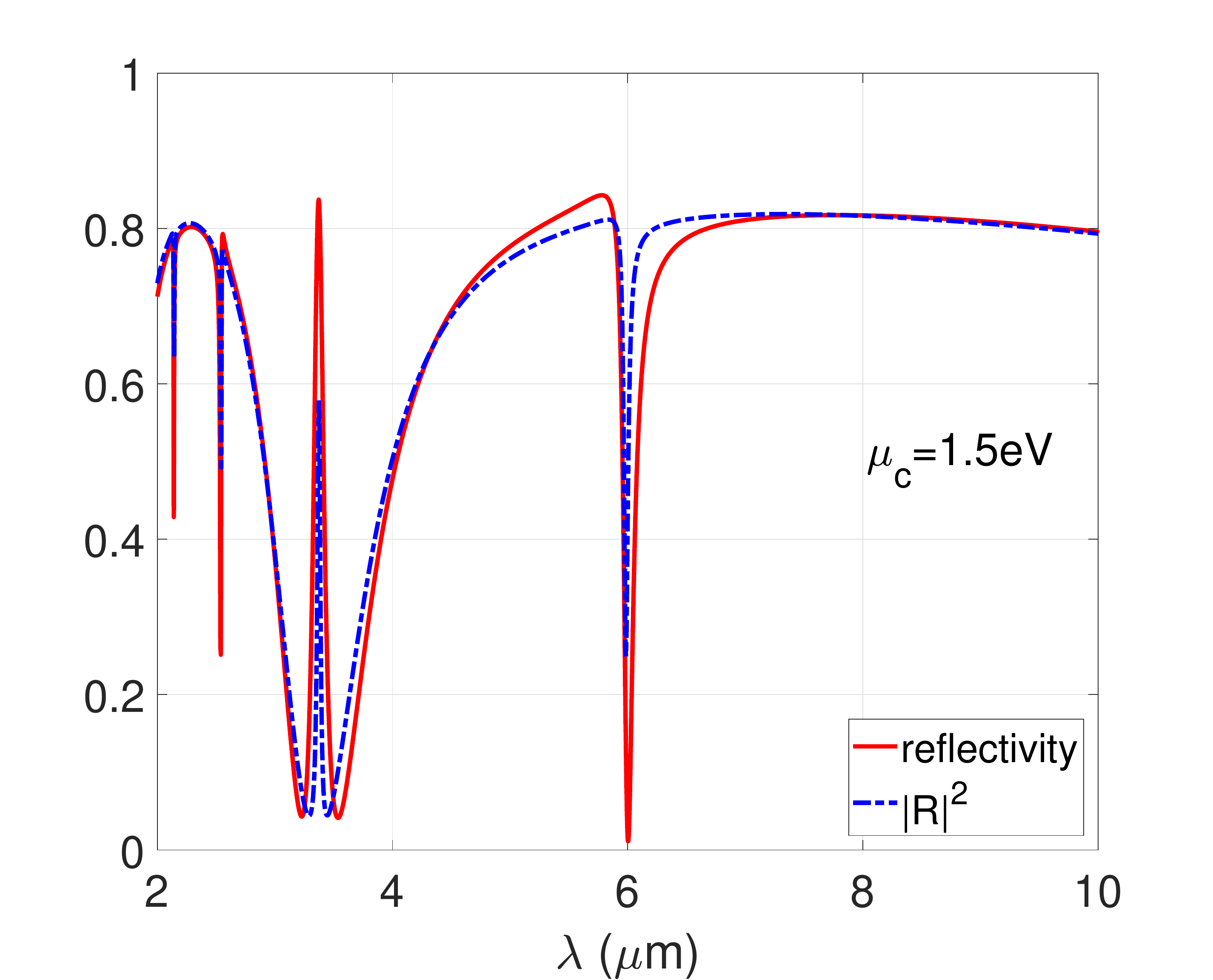}}
\centering  
\subfigure [\label{compare_trans_T_muc_1500_sp_10nm} transmission for $\mu_c=1.5eV$]
{\includegraphics[width=0.49\textwidth]{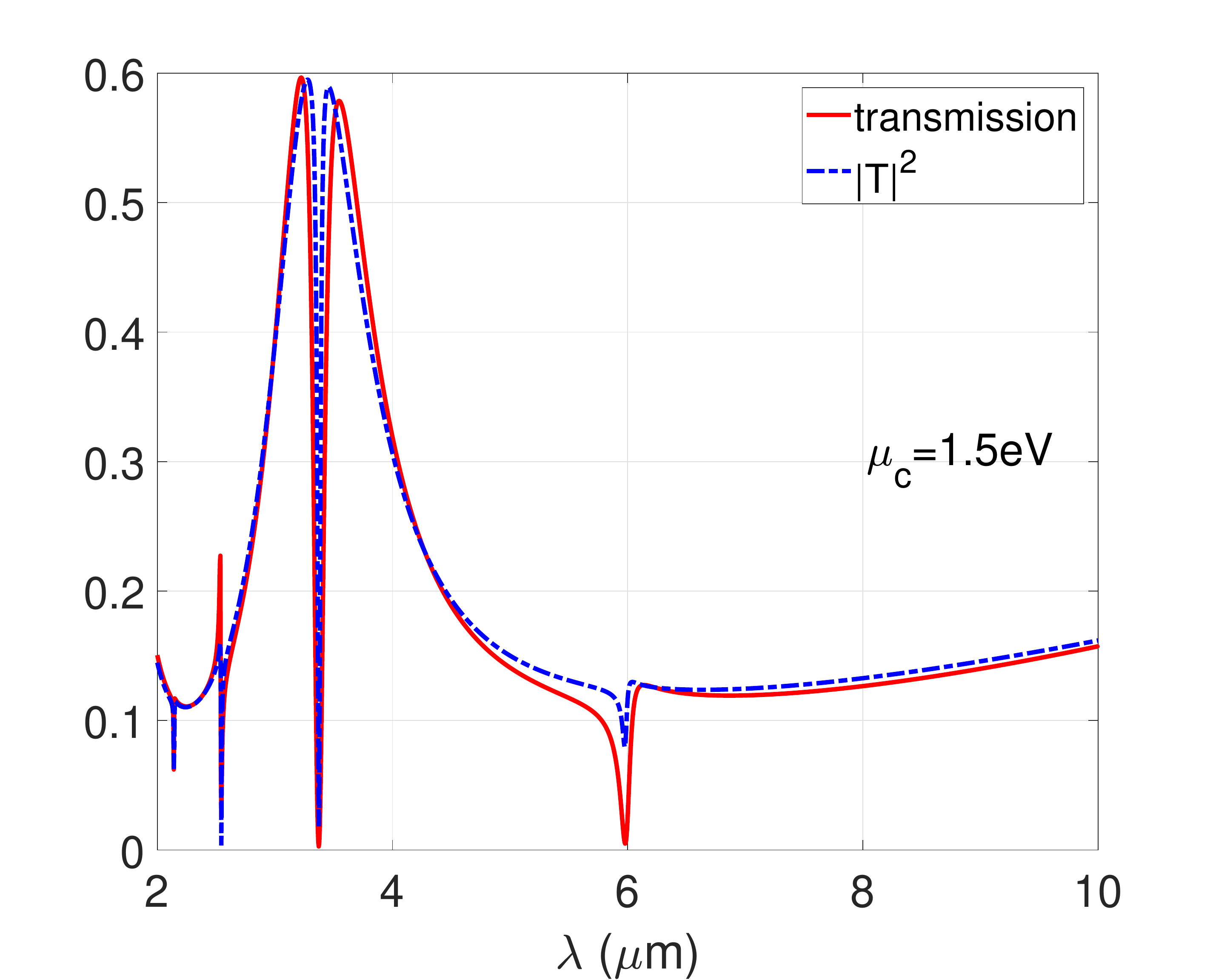}}
\caption{Comparison between the spectra of the hybrid-structure  with the reflection and transmission curves obtained from the PMM for two values of the chemical potential $\mu_c$. The chemical potential  is set to $\mu_c=1eV$, in Figs. (\ref{compare_reflec_R_muc_1000_sp_10nm}) and (\ref{compare_trans_T_muc_1000_sp_10nm}), while  $\mu_c=1.5eV$, in Figs. (\ref{compare_reflec_R_muc_1500_sp_10nm}) and (\ref{compare_trans_T_muc_1500_sp_10nm}). All these results fit very well with the rigorous numerical simulations obtained with the PMM.  Our model captures very well all resonances occurring in the hybrid system namely  Lorentz and  Fano ones. Parameters:  $\lambda \in [2,10]\mu m$, $\varepsilon^{(1)}=\varepsilon^{(3)}=\varepsilon^{(s)}=1$, $\varepsilon^{(2)}=1.54^2$, incidence angle= $0^{o}$.}
\end{figure}

Consider now the strongly coupled system sketched in Figs. (\ref{sub_system_2a}) and (\ref{sub_system_2b}). Since  the transverse geometrical parameters of the grating are smaller than the incident field wavelength $\lambda$ ($d<<\lambda$), 
we can introduce for the lattice mode an effective index $\alpha_0^{(1)}$ along the $x$-axis as follows: 
\begin{equation}
\alpha_0^{(1)}=\sqrt{\varepsilon^{(1)}-\gamma_0^{(1)2}}, 
\end{equation}
where $\alpha_0^{(1)}$ has a positive real part and a negative imaginary part. The $S$-parameters of the equivalent two ports network of Fig. (\ref{sub_system_2a}) are then given by :
\begin{figure}[!ht]
\centering 
{\includegraphics[width=0.6\textwidth]{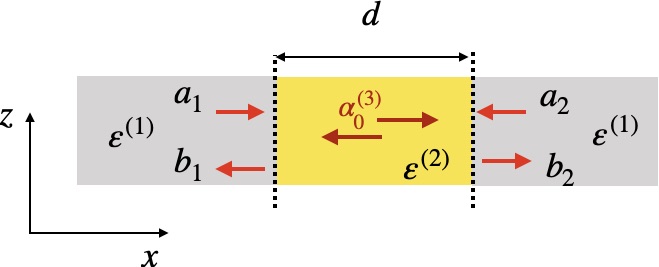}}
\caption{\label{sub_system_2a} Sketch showing the strong coupling between the gap plasmon mode $\alpha_0^{(3)}$ living in an $\sqrt{\varepsilon^{(2)}}$  homogeneous medium and $\alpha_0^{(1)}$ lattice mode in an $\sqrt{\varepsilon^{(1)}}$ effective homogeneous medium.}
\end{figure}
\begin{equation}\label{parametre_S}
\begin{bmatrix}
S_{11} & S_{12}\\
S_{21} & S_{22}
\end{bmatrix}
\begin{bmatrix}
a_1\\
a_2
\end{bmatrix}
=
\begin{bmatrix}
b_1\\
b_2
\end{bmatrix}
\end{equation}
where
\begin{equation}\label{Parametres2}
\left\{\begin{array}{l}
S_{11}(\omega)=S_{22}(\omega)=\dfrac{\left[1-n^2(\omega)\right]\left[1-\phi^2(\omega)\right]}{\left[1+n(\omega)\right]^2-\left[1-n(\omega)\right]^2 \phi^2(\omega)} \\[1em]
S_{12}(\omega)=S_{21}(\omega)=\dfrac{4n(\omega)\phi(\omega)}{\left[1+n(\omega)\right]^2-\left[1-n(\omega)\right]^2 \phi^2(\omega)}
\end{array}\right.,
\end{equation}
with
\begin{equation}
\left\{\begin{array}{l}
n(\omega)=\dfrac{\alpha_0^{(3)}(\omega)/\varepsilon^{(3)}(\omega)}{\alpha_0^{(1)}(\omega)/\varepsilon^{(1)}(\omega)}\\[1em]
\phi=e^{-i k_0\alpha_0^{(3)}d}.
\end{array}\right.,
\end{equation}
The dispersion relation of  this system is obtained by finding the zeros of the determinant $\Delta(\omega)$ of the matrix $S(\omega)$ of equation Eq. (\ref{parametre_S}) : 
\begin{equation}\label{Delta}
\Delta(\omega)=S_{11}(\omega)S_{22}(\omega)-S_{12}(\omega)S_{21}(\omega)= \left[S_{11}(\omega)-S_{12}(\omega)\right]\left[S_{11}(\omega)+S_{12}(\omega)\right]=0.
\end{equation}
Then we have two classes of solutions:
\begin{equation}\label{Delta4}
\left\{\begin{array}{l}
S_{11}(\omega)-S_{12}(\omega)=0\\
\text{or}\\
S_{11}(\omega)+S_{12}(\omega)=0 
\end{array}\right..
\end{equation}
As shown Fig. (\ref{compare_reflec_13_muc_1000_sp_10nm}) the resonance frequencies defined by the class of solutions satisfying to $S_{11}(\omega)+S_{12}(\omega)=0$ match with the narrow band resonances of the hybrid structure. Let us  set  
\begin{equation}
r_{13}(\omega)=-\left(S_{11}(\omega)+S_{12}(\omega)\right).
\end{equation}
Coefficient $r_{13}$ corresponds to the reflection coefficient of the strongly coupled system where the output and input ports are excited by two fields of equal amplitudes $a_1=a_2$. Therefore the reflection spectrum of the whole system can take the following form :
\begin{equation}\label{reflection approximation}
R=\dfrac{r_1+\phi_1 r_{13}r_2 \phi_2}{1+r_1\phi_1 r_{13}r_2 \phi_2}    
\end{equation}
and
\begin{equation}\label{transmission approximation}
T=\dfrac{t_1r_{13}t_2 \phi_2}{1+r_1\phi_1 r_{13} r_2 \phi_2}.
\end{equation}
By using the approximate model of Eqs. (\ref{reflection approximation}) and (\ref{transmission approximation}), we provide some numerical simulations (In Figs. (\ref{compare_reflec_R_muc_1000_sp_10nm}), (\ref{compare_trans_T_muc_1000_sp_10nm}), (\ref{compare_reflec_R_muc_1500_sp_10nm}) and (\ref{compare_trans_T_muc_1500_sp_10nm})) for different values of $\mu_c$. In these figures, we compare the spectra of the hybrid-structure  with the reflection and transmission curves obtained from rigorous PMM computations. The chemical potential is set to $\mu_c=1eV$, in Figs. (\ref{compare_reflec_R_muc_1000_sp_10nm}) and (\ref{compare_trans_T_muc_1000_sp_10nm}), while  $\mu_c=1.5eV$, in Figs. (\ref{compare_reflec_R_muc_1500_sp_10nm}) and (\ref{compare_trans_T_muc_1500_sp_10nm}). All these results fit very well the rigorous numerical simulations obtained with the PMM.  Our model captures very well all resonances phenomena occurring in the hybrid system namely  Lorentz and  Fano resonances  and thus confirms that couplings between some fundamental modes of  elementary sub-structures are of fundamental importance in these phenomena.
Armed with this model, we are now ready to deepen the explanation of the dispersion curves  of Figs. (\ref{spectrum_muc_1000_sp_10nm}) and (\ref{spectrum_muc_1500_sp_10nm}).
\section{Analysis of the Lorentz and Fano resonances of the system}  
\begin{figure}[h]
\centering  
\subfigure [\label{real_alpha_3_spacer}$\mu_c=1eV$]
{\includegraphics[width=0.49\textwidth]{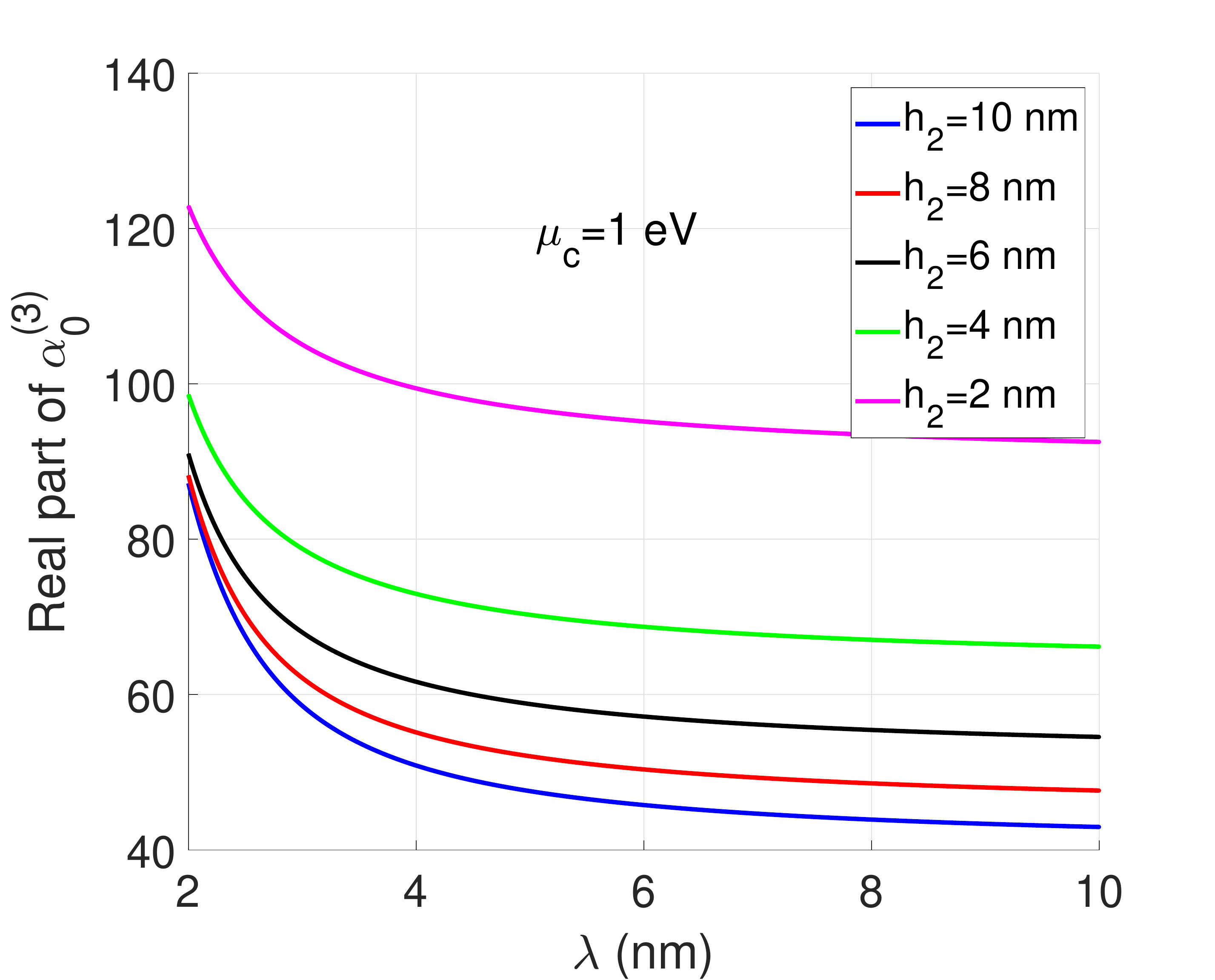}}
\centering  
\subfigure [\label{real_alpha_3_muc} $h_2=10 nm$]
{\includegraphics[width=0.49\textwidth]{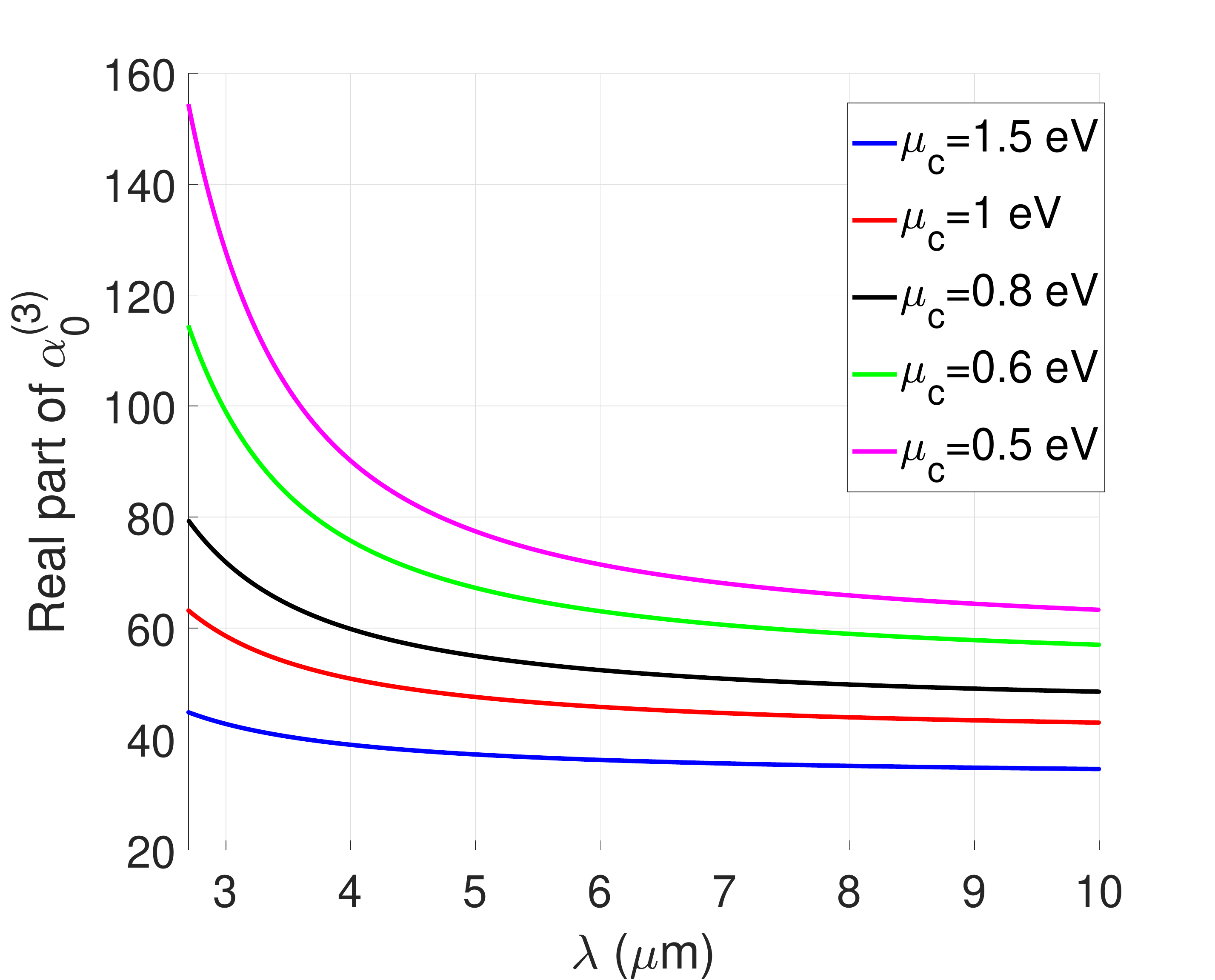}}
\caption{Dispersion curves of the effective index  $\alpha_0^{(3)}$ for different values of  $h_2$, ($\mu_c=1eV$) (Fig. (\ref{real_alpha_3_spacer})) and for different values of $\mu_c$ ($h_2=10 nm$) (Fig. (\ref{real_alpha_3_muc})). Increasing the chemical potential $\mu_c$ or the spacer width $h_2$, the real part of $\alpha_0^{(3)}$  decreases. Parameters: $\varepsilon^{2}=1.54^2$.}
\end{figure}

Analysing the reflection $|R|^2$, transmission $|T|^2$ from Eqs. (\ref{reflection approximation}), (\ref{transmission approximation}), it is possible to provide justifications for the curves shapes in Figs. (\ref{spectrum_muc_1000_sp_10nm}) and (\ref{spectrum_muc_1500_sp_10nm}). From these figures, we remark that:
\begin{enumerate}
    \item In the frequency range close to the resonance frequencies of the weak sub-system, the reflection and transmission spectra generally exhibit asymmetric Fano-like shapes while the absorption presents Lorentz-like shapes (left inserts of Figs. (\ref{spectrum_muc_1000_sp_10nm}) and (\ref{spectrum_muc_1500_sp_10nm})).
    \item When the resonance frequency of both  strongly and weakly coupled systems match each other, it results in an exaltation of the reflection and annihilation of both  transmission and absorption. This can be seen as a sort of induced reflection.   
    \item In the frequency range far from the resonance frequencies of the weakly coupled sub-system, a Lorentz-like  absorption enhancement can be observed (right inserts of Figs. (\ref{spectrum_muc_1000_sp_10nm}) and (\ref{spectrum_muc_1500_sp_10nm})). The scattering efficiency vanishes and the absorption takes its maximum value close to unity.
 \end{enumerate}
Before commenting on the first  point raised above, let us recall that, in general, the Fano resonance occurs when a narrow band resonance sub-system interferes with a continuum or a broadband resonance sub-system. The signature of this resonance in  the spectrum is the presence of two closed critical points corresponding to a vanishing value of the amplitude followed or preceded by an enhancement. In the current case, the zeros of the transmission $T$, in Eq. (\ref{transmission approximation}), are the zeros of the coefficient $r_{13}$ and these frequency values are always followed or preceded by great or little transmission  enhancements. Therefore the Fano resonance shape becomes obvious.\\  
For the second point, let us recall the resonance condition of the first sub-system. It is obtained from the  zeros of the reflection coefficient $R_{12}$, in Eq. \ref{reflection}, as soon as:
\begin{eqnarray}\label{resonance1}
\phi_1 r_2 \phi_3 \simeq -r_1 \text{ and }
1+r_1\phi_1 r_2 \phi_2 \neq 0 , 
\end{eqnarray} 
There is an extinction of the reflection without any annihilation of the transmission. Knowing that the resonance condition of the strongly coupled system is given by 
\begin{equation}\label{resonance2}
r_{13}(\omega) \simeq 0,
\end{equation}
when the latter resonance condition Eq. (\ref{resonance2}) meets the former Eq. (\ref{resonance1}), it results
\begin{equation}\label{resonance3}
r_{13}\phi_1 r_2 \phi_3 \simeq 0
\end{equation}
which leads to an exaltation of reflection, and an annihilation of the  transmission ( see Eq. \ref{transmission approximation}) and the absorption. The spectral responses of the structure  are shown to be highly tunable by changing a gate voltage applied to the graphene sheet.
{{The height $h_2$ of the horizontal cavity influences the system through the effective index $\alpha_0^{(3)}$. The dispersion curves of the effective index  $\alpha_0^{(3)}$ are plotted in Fig. \ref{real_alpha_3_spacer} for different values of $h_2$  while $\mu_c$  is kept constant and equal to $1eV$. It can be seen that increasing $h_2$ leads to a decrease of the real part of $\alpha_0^{(3)}$.
Since the $x$ dependance of the electromagnetic field  in the cavity may be approximated by $H_y(x)=A^+ exp(ik \alpha_0^{(3)}x)+A^- exp(-ik \alpha_0^{(3)}x)$, ($k=2\pi/\lambda$),  for a given $d$-length cavity,  the resonance wavelengths can be approximately obtained  through  a phase condition on the term $A^{\pm} sin(2\pi d \alpha_0^{(3)}/\lambda_r$). When $\alpha_0^{(3)}$ decreases, the resonance wavelength $\lambda_r$ brought by the strongly coupled  sub-system also decreases. Consequently increasing spacer height pushes the resonance wavelengths  resulting from the strongly coupling sub-system towards the visible wavelengths range. The same behavior can be observed when the height $h_2$ is kept constant while increasing the chemical potential $\mu_c$ (see figure \ref{real_alpha_3_muc}). This time it is $\mu_c$ that influences the system through the effective index of the horizontal cavity. Increasing $\mu_c$ decreases $\alpha_0^{(3)}$ and thereby leads to a decrease of the resonance wavelengths.}}   
By tuning the potential $\mu_c$, one can realize the condition of Eq. (\ref{resonance3}) leading to an induced reflection phenomenon. 
For the last point raised, the Lorentz resonance shape of the absorption is provided by the poles of the scattering parameters of the system \textsl{i.e.} when $1+r_1\phi_1 r_{13} r_2 \phi_2 \simeq 0$ leading to weak values of both reflection and transmission. Besides, the exaltation of the absorption always occurs around frequencies where both reflection and transmission are weak and equal and these frequencies are different from the zeros of the coefficient $r_{13}$. Therefore in the frequency range far from the resonance frequency of the EOT sub-system, the hybrid structure behaves as a tunable perfect absorber. \\
{{Finally, it is worth noticing that the present model works very well for normal incidence and reasonably well for angles of incidence up to twenty degrees. For large angles of incidence, some new resonances appear in the spectra  and are not captured by our model.}}
\section{Conclusion}
{{In conclusion, we have proposed a simple model, allowing to deepen the comprehension of the resonance phenomena involving  the EOT phenomenon  and a metal/insulator/graphene gap plasmon excitation. We consider a hybrid structure that consists of a 1D array of periodic subwavelength slits ended by a metal/insulator/graphene gap. For our analysis, this hybrid structure is split into two sub-systems. Each sub-system is driven by eigenmodes operating in an appropriate coupling regime. The study of the first sub-system, characterised by modes operating in a weak coupling regime, allows to understand the broadband resonance of the hybrid system. We provided an analytical expression of the reflection and transmission coefficients of this first sub-system. The behavior of the second sub-system, characterized by modes acting in a regime of strong coupling allows to understand the narrow-band nature of the hybrid system. Here, the resonance frequencies directly depend on the metal-insulator-graphene horizontal Perot Fabry cavity effective index. Since the  real part of this  effective index decreases by increasing the graphene sheet chemical potential,   the resonance wavelengths of the system become perfectly tunable ; better yet an induced reflection phenomenon or  perfect absorption can be achieved with suited values of the graphene sheet Fermi level. We proposed a spectral function allowing not only to characterize the resonance frequencies of this second sub-system, but also showed that introducing this spectral function into the reflection and transmission coefficients of the first sub-system, we obtain an analytical expression of the reflection and transmission coefficients of the global hybrid system which are successfully compared with those obtained with rigorous numerical simulations (through the PMM approach). Finally, armed with these analytical expressions, we provided a full description  of the resonance phenomena occurring in  the system.
Our analysis in terms of simple modes couplings can be extended to study the coupling of the lattice modes with a substrate made by a non-reciprocal photonic topological materials, of particular interest for energy management and transport \cite{Doyeux} and for atomic manipulation \cite{Silveirinha}. The analysis of such complex hybrid configurations involving diffraction gratings coupled to hybrid graphene multilayer structures could also be applied to study and to estimate more complicated phenomena, like the Casimir effect \cite{Messina1} and the radiative heat transfer \cite{BZhao}.\\ }}
{\bf{Funding}}\\
This work has been sponsored by the French government research program "Investissements d'Avenir" through the IDEX-ISITE initiative 16-IDEX-0001 (CAP 20-25)\\

\end{document}